\documentclass[runningheads]{llncs}
\usepackage[T1]{fontenc}
\usepackage{graphicx}
\usepackage{amsmath}
\begin{document}
\title{Predicting Exoplanetary Features with a Residual Model for Uniform and Gaussian Distributions}
\titlerunning{Predicting Exoplanetary Features with RUG}
%
\author{Andrew Sweet}
%
\authorrunning{A. Sweet\inst{1}}
%
\institute{Assemi Group, Inc., Fresno CA 93704, USA
\email{asweet@assemigroup.com}}
%
\maketitle              
\begin{abstract}
The advancement of technology has led to rampant growth in data collection across almost every field, including astrophysics, with researchers turning to machine learning to process and analyze this data. One prominent example of this data in astrophysics is the atmospheric retrievals of exoplanets. In order to help bridge the gap between machine learning and astrophysics domain experts, the 2023 Ariel Data Challenge\footnote{https://www.ariel-datachallenge.space/} was hosted to predict posterior distributions of 7 exoplanetary features. The procedure outlined in this paper leveraged a combination of two deep learning models to address this challenge: a Multivariate Gaussian model that generates the mean and covariance matrix of a multivariate Gaussian distribution, and a Uniform Quantile model that predicts quantiles for use as the upper and lower bounds of a uniform distribution. Training of the Multivariate Gaussian model was found to be unstable, while training of the Uniform Quantile model was stable. An ensemble of uniform distributions was found to have competitive results during testing (posterior score of 696.43), and when combined with a multivariate Gaussian distribution achieved a final rank of third in the 2023 Ariel Data Challenge (final score of 681.57).
\end{abstract}
\section{Introduction}

\subsection{Motivation}
As the advent of big data in astrophysics continues to expand, machine learning is being adopted and further developed\footnote{https://ml4astro.github.io/icml2022/} for use in this field~\cite{szab_2022,VanderPlas_2012}, in particular for the computational efficiency and potentially improved robustness over classical techniques. Examples of applications include anomaly detection of rare supernova from photometric data~\cite{10.1093/mnras/stz2362}, classification of variable stars from light curves~\cite{Szklenar_2020}, and cosmic web simulations with generative adversarial networks (GANs)~\cite{Rodr_guez_2018}. One exciting area in this domain is analyzing the atmospheres of distant planets, called extra-solar planets or exoplanets, from spectroscopic data. With more satellites going up with increased capabilities for collecting this data, such as the James Web Space Telescope (JWST)\footnote{https://webb.nasa.gov/} launched in 2021, the Ariel Space Mission\footnote{https://arielmission.space/} expected to launch in 2029, and the Twinkle Space Telescope~\cite{Edwards_2018} expected to launch in 2024, further methods for analyzing and extracting meaningful insight in a timely and robust manner are needed. In order to address these needs and encourage collaboration across the domains of machine learning and astrophysics, Yip et al.~\cite{changeat2023esaariel,yip2022esaariel} created a simulated spectroscopic dataset and hosted the Ariel Data Challenge at the Conference and Workshop on Neural Information Processing Systems (NeurIPS) 2022 and at the European Conference on Machine Learning and Principles and Practice of Knowledge Discovery in Databases (ECML PPKD) 2023. The goal of these challenges is to predict posterior distributions of a number of atmospheric components and planetary features from spectroscopic readings and a few planetary system measurements. 

\subsection{Data}
\begin{figure}
\includegraphics[width=\textwidth]{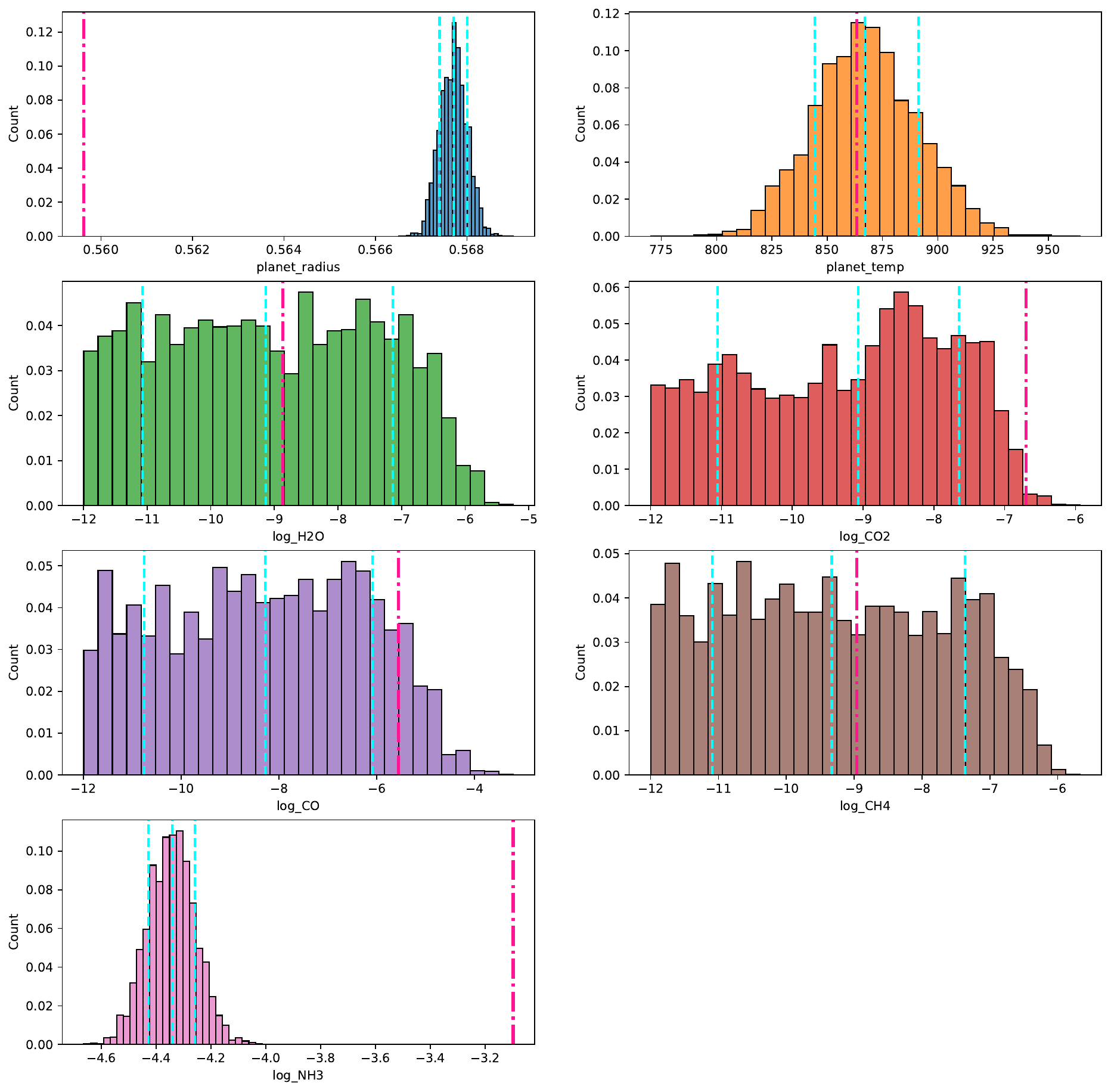}
\caption{Weighted histograms of the trace data for one planet. The dashed cyan lines represent the 16\textsuperscript{th}, 50\textsuperscript{th} and 84\textsuperscript{th} percentiles, and the dashed-dotted deep pink lines represent the forward model parameters. This illustrates the varied distribution shapes of the target values, and the forward model parameters not necessarily lining up with the 50\textsuperscript{th} percentile.} \label{output_data_sample}
\end{figure}

The Ariel Data Challenge at NeurIPS 2022 and ECML PPKD 2023 had three main differences: the size of the provided data sets, the number of exoplanetary features being predicted, and the scoring metrics. The methods here will focus on the data for ECML PPKD 2023\footnote{See Yip et al.~\cite{yip2022esaariel}. and Changeat et al.~\cite{changeat2023esaariel} for complete descriptions of the data and data generation processes.}. There were five data files distributed for the challenge, which can broadly be separated into input and output data for machine learning purposes, and contain simulated data for 41,423 (denoted as $N$ below) planets. For input data there were two files: 
\begin{itemize}
\item spectral data: which was composed of the wavelength grid, spectrum, uncertainty and bin width across 52 wavelength channels, of shape $N \times 4 \times 52$.
\item auxiliary data: containing 8 auxiliary features for each planetary system, of shape $N \times 8$. These features were the planet's mass, orbital period, radius, semi-major axis, and surface gravity, as well as it's host star's mass, radius, temperature, and distance from Earth.
\end{itemize}
The output data, also labelled as the ground truth, targets 7 values for each planet: the planet's radius, temperature, and 5 atmospheric readings, $log(H_2 O)$, $log(CO_2)$, $log(CH_4)$, $log(CO)$ and $log(NH_3)$. There were three files of output data:
\begin{itemize}
\item forward model parameter data: the inputs for each target value given to the forward model used in the data simulation process, of shape $N \times 7$, and is available for all $N$ planets in the training data.
\item trace data: contains the posterior distributions from MultiNest of the 7 target values given to the forward model, and associated importance weights per sample for 6766 ($N_T$) planets.
\item quartiles data: the 16\textsuperscript{th}, 50\textsuperscript{th} and 84\textsuperscript{th} percentiles estimated for each of the 7 targets, given with shape $N_T \times 3 \times 7$. This data is only available for the same 6766 planets as in the trace data.
\end{itemize}
Figure~\ref{output_data_sample} shows an example of the values from the output data for one planet.

\section{Methodology}
The ultimate goal addressed here is to generate posterior distributions of 7 planetary features, given the input spectra and auxiliary features of the planets. Three models were used, with the first one as a pre-training step to help the subsequent training phases converge and the latter two used to generate the posterior distributions. The following subsections describe the implementation process in greater detail and all code is available on GitHub\footnote{https://github.com/acsweet/ariel\_data\_challenge\_2023}.

\subsection{Models}
There were three models used in this setup, and all were built with Keras~\cite{chollet2015keras} using TensorFlow~\cite{tensorflow2015-whitepaper}.\footnote{Assume any parameters not specified are defaults.} Each model shared the same backbone, and differed only in the final layer(s), also called heads. The loss functions used are discussed with each of the heads in the following sections.

\subsubsection{Model Backbone}
\begin{figure}
\centering
\includegraphics[height=7cm]{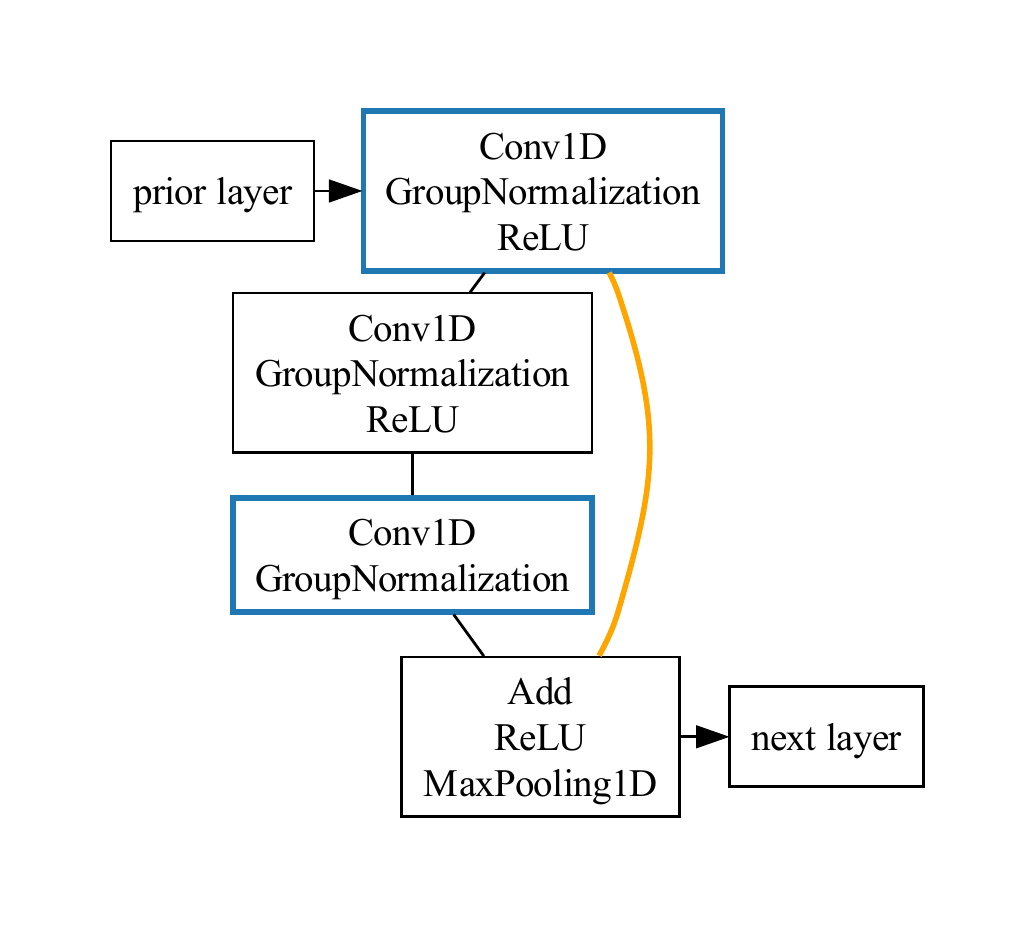}
\caption{Resnet block with a skip connection shown in orange. The output of the two blue nodes is added in the final node, resulting in the residual skip connection.} \label{resnetBlockGraph}
\end{figure}

The baseline model\footnote{https://github.com/ucl-exoplanets/ADC2023-baseline} provided for the challenge consisted of a convolutional neural network (CNN) adapted from Yip et al.~\cite{yip2021peeking}. Specifically, the baseline model used 1D convolutional layers applied to the spectral inputs. In order to leverage the strength of these layers and allow for a deeper network with more learnable parameters, or weights, residual blocks with skip connections were created, inspired by Residual Networks (ResNets)~\cite{he2015deep}. These skip connections act as a kind of regularization, allowing the network to go deeper while mitigating overfitting. Figure~\ref{resnetBlockGraph} outlines the construction of a residual block, where three 1D convolutions were performed, each followed by a group normalization layer~\cite{wu2018group}. All of the group normalizations were followed by a ReLU (rectified linear unit) activation function, except for the last which applied the activation after the addition with the skip connection. Finally a 1D max pooling with a window size of 2 was applied to reduce the number of channels in half. The residual blocks and parameters are shown in Table~\ref{resnetBlockSummary}, with the difference between each block being the number of kernels, $k$, used per block. From the first residual block to the last, these were 16, 32, 64 and 128 kernels. The auxiliary feature input was followed by a single dense layer and then concatenated with a flattened output of the final residual block. Then two subsequent dense layers, each followed by a dropout layer, were applied to the concatenated layer's output. Each dense layer had a ReLU activation function applied to its output. An overview of the backbone portion of the models can be seen in Table~\ref{resnet_backbone}. 

\begin{table}
\centering
\caption{Residual Block setup and parameters. The parameters are defined as follows: $m$ is the batch size, $s$ is the steps or number of input features, $c$ represents the number of channels, $ks$ is the kernel size, $p$ is the pooling window size, $g$ is the number of groups, and $k$ is the number of kernels assigned to this block. With this setup $c_i$ will be equal to $k$ and $c_{i-1}$ is the number of channels from the prior layer. The inputs to the add layer are the outputs of pre\_act and gn\_2.}
\label{resnetBlockSummary}
\begin{tabular}{lclc}
\hline
Layer (type) & Configuration & Output Shape & Weights \\
\hline
prior layer & & ($m$, s, $c_{i-1}$) & 0 \\
 pre\_conv (Conv1D) & $ks=3$ & ($m$, s, $c_i$) & $k*ks*c_{i-1} + k$ \\
 pre\_gn (GroupNorm) & $g=1$ & ($m$, s, $c_i$) & $2*c_i$ \\
 pre\_act (ReLU) & & ($m$, s, $c_i$) & 0 \\
 rb\_conv\_1 (Conv1D) & $ks=3$ & ($m$, s, $c_i$) & $k*ks*c_i + k$ \\
 gn\_1 (GroupNorm) & $g=1$ & ($m$, s, $c_i$) &  $2*c_i$ \\
 act\_1 (ReLU) & & ($m$, s, $c_i$) & 0 \\
 rb\_conv\_2 (Conv1D) & $ks=3$ & ($m$, s, $c_i$) & $k*ks*c_i + k$ \\
 gn\_2 (GroupNorm) & $g=1$ & ($m$, s, $c_i$) & $2*c_i$ \\
 add (Add) & & ($m$, s, $c_i$) & 0 \\
 post\_add\_act (ReLU) & & ($m$, s, $c_i$) & 0 \\
 maxpool (MaxPooling1D) & $p=2$ & ($m$, $\lfloor s/2 \rfloor$, $c_i$) & 0 \\
\hline
\end{tabular}
\end{table}

\begin{table}
\centering
\caption{Backbone used in all models. In the table below, $m$ represents the batch size, $k$ the number of kernels, and $p$ the dropout probability. The concatenate layer combines the outputs of the flatten and aux\_dense\_1 layers.}
\label{resnet_backbone}
\begin{tabular}{lclc}
\hline
Layer (type) & Configuration & Output Shape & Weights \\
\hline
 spectra\_input (InputLayer) & & ($m$, 52, 1) & 0 \\
 res\_block\_0 (ResidualBlock) & $k=16$ & ($m$, 26, 16) & 1728 \\
 res\_block\_1 (ResidualBlock) & $k=32$ & ($m$, 13, 32) & 7968 \\
 res\_block\_2 (ResidualBlock) & $k=64$ & ($m$, 6, 64) & 31296 \\
 res\_block\_3 (ResidualBlock) & $k=128$ & ($m$, 3, 128) & 124032 \\
 auxiliary\_input (InputLayer) & & ($m$, 17) & 0 \\
 flatten (Flatten) & & ($m$, 384) & 0 \\
 aux\_dense\_1 (Dense) & & ($m$, 500) & 8500 \\
 concatenate (Concatenate) & & ($m$, 884) & 0 \\
 concat\_dense\_1 (Dense) & & ($m$, 500) & 442500 \\
 dropout\_1 (Dropout) & $p=0.1$ & ($m$, 500) & 0 \\
 concat\_dense\_2 (Dense) & & ($m$, 100) & 50100 \\
 dropout\_2 (Dropout) & $p=0.1$ & ($m$, 100) & 0 \\
 \hline
 Total trainable weights & & & 666124 \\
\hline
\end{tabular}
\end{table}

\subsubsection{Pretraining Model} 
The first model was used for pre-training for the subsequent two models. It started with the backbone model shown in Table~\ref{resnet_backbone}, with an attached output head of a simple dense fully connected layer with 7 units, equal to the number of target features, and no activation function. The model was trained using the mean squared error metric (MSE) as the loss function, which is defined as the mean of the squared difference between the true, $y$, and predicted, $\hat{y}$, values and shown in Equation~\ref{mse}. A graph of the Pretraining model can be seen in Figure~\ref{phase1model} in the Appendix.

\begin{equation}
\mathcal{L}(y, \hat{y}) = \frac{1}{n}\sum_{i=1}^{n}(y_i - \hat{y}_i)^2
\label{mse}
\end{equation}

\subsubsection{Multivariate Gaussian Model}
After pre-training, the heads were swapped and a new learning objective with a different loss function was set. The first of these two models outputs parameters for a multivariate Gaussian distribution, based on the work of Mart{\'{\i}}nez et al.~\cite{Ard_vol_Mart_nez_2022}. These distributions are defined with a mean vector, $\mu$, and covariance matrix, $\Sigma$. The covariance matrix can be represented by its Cholesky decomposition, namely $\Sigma = (LL^T)^{-1}$, where $L$ is a lower triangular matrix, resulting in the loss function from Cobb et al.~\cite{cobb_2019}, shown in Equation~\ref{chol_mult_gauss_loss}. 

%

\begin{equation}
\mathcal{L}(y, \mu, L) = -2 \sum_{d=1}^{D} log(l_{dd})+(y-\mu)^T LL^T (y-\mu)
\label{chol_mult_gauss_loss}
\end{equation}

The covariance matrix, $\Sigma$, and subsequent $L$ matrices are square and of order $D$. In this setting, $D$ is equal to the number of target features, and $l_{dd}$ represents the $d^{th}$ diagonal element of $L$. The output head of this model was two separate fully connected dense layers. One represented $\mu$ with 7 units and a sigmoid activation function. And the other had $D (D + 1) / 2$ units with a linear activation, where $D = 7$, and represented the elements of the lower triangular matrix $L$ flattened into a vector. The output corresponding to the covariance matrix was reshaped into a lower triangular matrix for use in the loss function and reconstructing the covariance matrix to use for sampling. During training and inference the exponential of the diagonal values, $l_{dd}$, of $L$ was taken so that $(LL)^T$ is positive-definite, as noted in~\cite{cobb_2019}. Figure~\ref{phase2MGmodel} in the Appendix displays a graph of this model.

\subsubsection{Uniform Quantile Model}
The second model after pre-training was setup to output the parameters needed for uniform distributions. This was done by predicting the 5\textsuperscript{th}, 50\textsuperscript{th} and 95\textsuperscript{th} percentiles, where the 5\textsuperscript{th} and 95\textsuperscript{th} percentiles could be used as the upper and lower bounds for a uniform distribution. These percentiles were chosen to capture the approximate boundaries of the distribution. 

The quantile loss function~\cite{10.1257/jep.15.4.143}, $\rho_{\tau}$, is defined in Equation~\ref{quantile_loss}, where $\tau$ represents the desired quantile. Similar to the methods outlined in~\cite{sherwood_2022}, the absolute value term, $|y - \hat{y}|$, in $\rho_{\tau}$ can be approximated with the Huber loss function~\cite{huber_1964} from Equation~\ref{huber_loss}. The Huber loss function, $h_{\delta}$, is defined with a parameter $\delta$ near 0, and allows $\rho_{\tau}$ to be smooth around $|y - \hat{y}|=0$. $y$ the target values sampled from the trace data, and $\hat{y}$ the quantile predictions.

\begin{equation}
\rho_{\tau}(y, \hat{y}) = \begin{cases}
		(1 - \tau) |y - \hat{y}|& \text{if } y \le \hat{y} \\
		\tau |y -\hat{y} | & \text{if } y > \hat{y}
	\end{cases}
\label{quantile_loss}
\end{equation}

\begin{equation}
h_{\delta}(y, \hat{y}) = \begin{cases}
		\frac{(y - \hat{y})^2}{2\delta} & \text{if } |y - \hat{y}| \le \delta \\
		(|y - \hat{y}| - \frac{\delta}{2}) & \text{otherwise} 
	\end{cases}
\label{huber_loss}
\end{equation}

Putting those two equations together, Equation~\ref{final_quantile_loss} shows the final loss function for the Uniform Quantile model with an additional regularization term, $\Lambda_\tau$, that returns the maximum of 0 and the difference between the prediction for the current quantile and the prediction for the max quantile, shown in Equation~\ref{quantile_regularization}. This effectively penalized the quantile predictions further if they were greater than what should have been the greatest quantile. An implementation vectorized to calculate the loss against multiple quantiles simultaneously can be seen in~\cite{biolatti_quantile}.


\begin{equation}
\mathcal{L}_{\tau, \delta}(y, \hat{y}, \hat{y}_{\tau_{max}}) = \begin{cases}
		(1 - \tau) h_{\delta}(y, \hat{y}) & \text{if } y \le \hat{y} \\
		\tau h_{\delta}(y, \hat{y})  & \text{if } y > \hat{y}
	\end{cases} + \Lambda_\tau(\hat{y}, \hat{y}_{\tau_{max}})
\label{final_quantile_loss}
\end{equation}

\begin{equation}
\Lambda_\tau(\hat{y}, \hat{y}_{\tau_{max}}) = max(\hat{y} - \hat{y}_{\tau_{max}}, 0)
\label{quantile_regularization}
\end{equation}

Here $\delta$ was assigned 0.0001. The target quantile values were set to 0.05, 0.5 and 0.95, and the model was declared with 7 fully connected dense layers each with 3 output units and no activation function, for the 7 target features and 3 target quantiles respectively. The 0.05 and 0.95 quantiles would be used as approximations for the lower and upper bounds of uniform distributions for sampling. A graph of the Uniform Quantile model is displayed in Figure~\ref{phase2UQmodel} of the Appendix.

\subsection{Data Processing}
Preprocessing was done on both the inputs and outputs of each of the models, primarily with the python packages NumPy~\cite{harris2020array} and scikit-learn~\cite{scikit-learn}. The inputs fall under two categories, the spectra and the auxiliary features. For the spectra, noise was added based on the provided uncertainty, sampled from a normal distribution with mean 0, and standard deviation equal to the spectra noise with the uncertainty values provided for each wavelength channel. Then the spectra were normalized with the mean of the spectrum values for each planet, similar to part of the preprocessing done in~\cite{Ard_vol_Mart_nez_2022}. For the auxiliary features, summary statistics of the raw spectral data were added, specifically the minimum, maximum, mean, and standard deviation. Additionally, four extra features were calculated: the ratio of the planet’s mass to its star’s mass, an approximate calculation of the planet’s equilibrium temperature, the planet’s host star’s density, and an approximate calculation of the planet’s semi-major axis\footnote{Note: the last calculation is similar to the planet\_distance feature included in the auxiliary data.}. This resulted in 16 auxiliary features input into the models. Equation~\ref{planet_equilibrium} shows the calculation of the planet's equilibrium temperature, where $T_{star}$ and $R_{star}$ represent the temperature and radius of the star, $a$ is the orbital distance of the planet or its semi-major axis, and $A$ represents an albedo constant. The term containing $(1 - A)$ was excluded and expected to be represented in the weights learned during training. See~\cite{Mendez_2017} for a good discussion on this topic, with two additional constants in the $(1 - A)$ term.

\begin{equation}
T_{equlibrium} = T_{star}(1 - A)^{\frac{1}{4}}\left(\frac{R_{star}}{2a}\right)^{\frac{1}{2}}
\label{planet_equilibrium}
\end{equation}

The planet's semi-major axis, $a$, was approximated with Equation~\ref{planet_semimajor_axis}, which is derived from Kepler's Third Law. Below, $O$ is the orbital period, $G$ is the gravitational constant, and $M_1$ and $M_2$ are the masses of body 1 and 2. The gravitational constant was excluded in this calculated feature as multiplication by a constant would not change the values after the subsequent normalization was applied.

\begin{equation}
a = \left(\left(\frac{O}{2\pi} \right)^2 G (M_1 + M_2) \right)^{\frac{1}{3}}
\label{planet_semimajor_axis}
\end{equation}

A natural log transform was done on the auxiliary features to account for the skew in the distributions of these features. Finally, they were normalized to have a mean of 0 and standard deviation of 1, using scikit-learn's StandardScaler. 

For the output features, min-max scaling was used, normalizing each feature between 0 and 1, as shown in Equation~\ref{min_max}, where $x$ represents a feature vector and $x'$ is the normalized feature. The default prior bounds of the targets in Table~\ref{default_priors} were provided and used as the min and max of each output feature.

\begin{equation}
x' = \frac{x - min(x)}{max(x) - min(x)}
\label{min_max}
\end{equation}

\begin{table}
\centering
\caption{Default prior bounds. All atmospheric gasses had the same bounds.}
\label{default_priors}
\begin{tabular}{lcc}
\hline
Feature & Lower Bound & Upper Bound \\
\hline
planet\_radius & 0.1 & 3 \\
planet\_temperature & 0 & 7000 \\
gasses & -12 & -1 \\
\hline
\end{tabular}
\end{table}

\subsection{Training}
All models were trained using a technique called K-fold cross validation~\cite{raschka2020model}, with scikit-learn's KFold module. K-fold cross validation involves randomly splitting the data into $K$ equal sized partitions. Note that a capital $K$ is used here to distinguish it from the lowercase $k$ used earlier to represent kernels. Training is then conducted $K$ times, where $K - 1$ partitions are used as a training set, and the last partition is held out to use as a validation set. To report on accuracy in the Results and Discussion, a test set of 10\% was held out before doing the K-fold partitioning, but all of the data was used in the final training done for the challenge. The ultimate goal of this style of training was to use the models trained on each fold in an ensemble prediction at the end. It has been shown that ensembles of models can perform better than individual models, in particular when there is diversity among the models~\cite{BROWN20055}. This diversity was expected to be stimulated by having each model trained on varied subsets of the data.

Training was conducted in two phases. Phase 1 trained the Pretraining model with the loss from Equation~\ref{mse}. It was conducted with a batch size of 1024 and 50 epochs, using the Adam optimizer~\cite{kingma2017adam} with a learning rate $\alpha = 0.001$. For each batch, the training data was sampled randomly and the spectral inputs were augmented with added noise as described in the Data Processing section. The training data with the augmented spectral inputs was repeated 10 times and the validation data was similarly repeated 5 times for each epoch. Early stopping was setup with the training with a patience of 3 epochs, such that if no improvement was made in the validation loss for 3 epochs, training was stopped and the best model weights were saved. A learning rate scheduler was also used, with a linearly increasing warmup from $\alpha = 0.0002$ to 0.001 in the first 5 epochs, then decaying over the remaining epochs to 0.00001.

Phase 2 of training involved separately training the Multivariate Gaussian model and the Uniform Quantile model. The inputs for both models were the same as in phase 1. The outputs for both the Multivariate Gaussian and Uniform Quantile models were generated similarly: for each batch, a single sample per planet of "true" values was drawn from the trace values provided weighted by the trace weights, which were then min-max normalized with the provided prior bounds. Thus, for both models the data was generated with noise added to the spectral inputs and a single weighted sample was drawn from the trace values as outputs for each planet. This process was repeated a number of times across all of the training and validation data for each epoch (to have a more meaningful loss metric reported at the end of each epoch.) For the Gaussian Multivariate model the training data was repeated 100 times and the validation data 25 times per epoch, and for the Uniform Quantile model the training data was repeated 30 times and validation data 5 times per epoch. Both models were set to train with a batch size of 1024 for 100 epochs, and set to stop early. The patience for early stopping was set to 10 epochs for the Gaussian Multivariate model and 4 epochs for the Uniform Quantile model. The Gaussian Multivariate model was trained with exactly the same learning rate scheduler and optimizer as the Pretraining model from Phase 1. The Uniform Quantile model was compiled with the Adam optimizer as well, but with a cyclical learning rate scheduler~\cite{smith2017cyclical}. This scheduler was declared with a max learning rate, $\alpha$, of 0.001, a min $\alpha$ of 0.00001 and declined exponentially at a factor of 0.95. 

%
%
%

\subsection{Inference and Evaluation}
The final output of these models is the posterior distributions of the 7 target features. For the Multivariate Gaussian model the outputs from the layer representing the lower triangular matrix was reshaped and processed back into the covariance matrix, $\Sigma$, and when combined with the outputs from the layer representing $\mu$ were used with NumPy's random.multivariate\_normal function to sample $N_{mg}$ values for each planet. With the Uniform Quantile model, the outputs representing the $5^{th}$ and $95^{th}$ percentiles were taken as the upper and lower bounds for a uniform distribution and sampled $N_{uq}$ times, with NumPy's random.uniform function. Final inference was then taken as an ensemble of multiple models, with an equal number of samples coming from each model in the ensemble.

The posterior distributions were evaluated on two metrics, a spectral score and a posterior score, with the $score_{final} = 0.2 \times score_{spectral} + 0.8 \times score_{posterior}$. The posterior score tested how similar the underlying distributions were. The spectral score tested the similarity between the combination of the uncertainty bounds and median spectra against spectra from Bayesian Nested Sampling~\cite{Al_Refaie_2021,Feroz_2009}. The scores were scaled to be between 0 and 1000, with 1000 being the highest and most similar. The calculations for the metrics were provided as part of the baseline~\cite{adc_baseline} and a more detailed explanation can be found in~\cite{adc_scoring}.

\section{Results and Discussion}
\begin{figure}[ht]
\includegraphics[width=\textwidth]{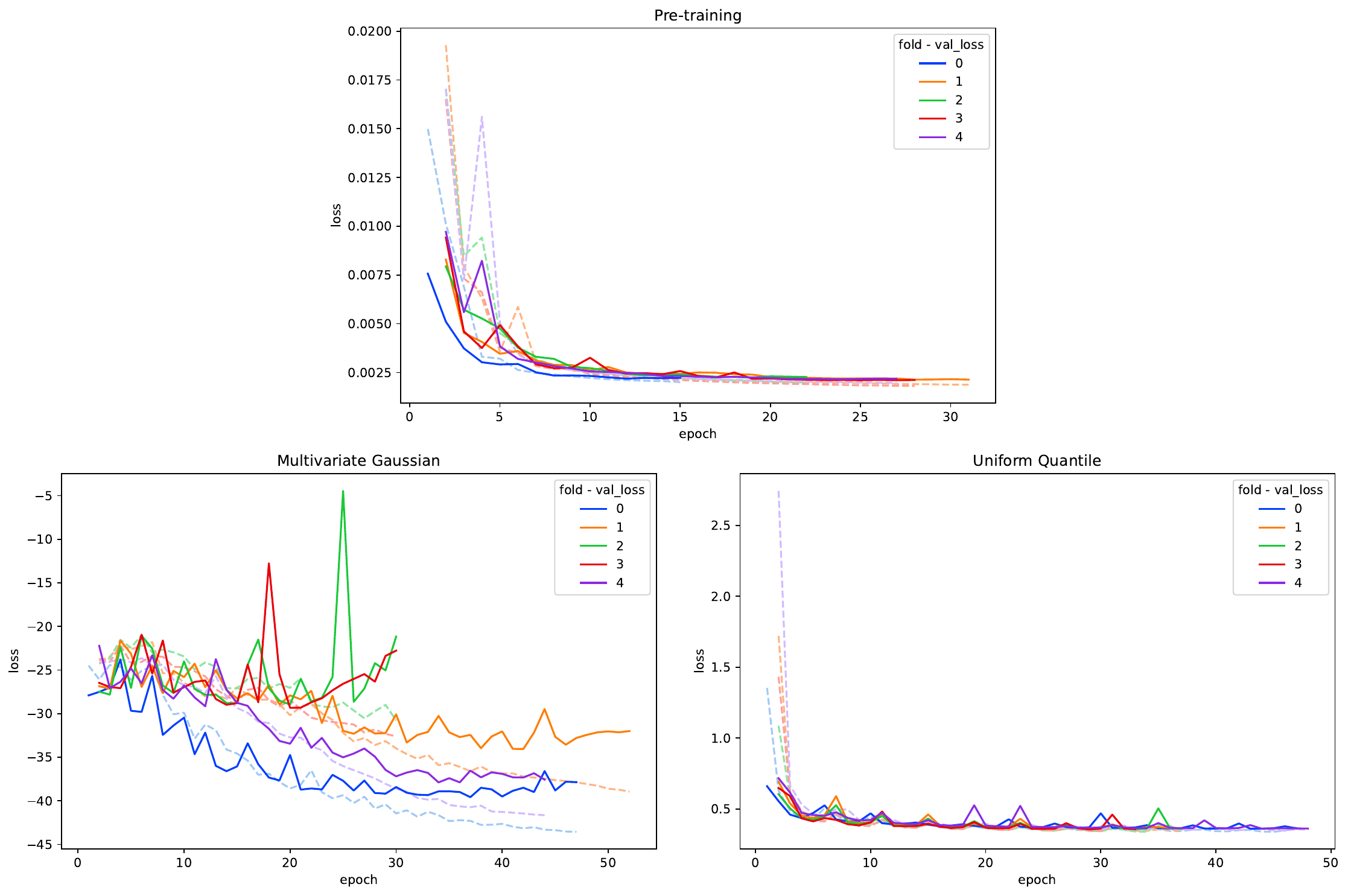}
\caption{Training loss. The solid lines represent the validation loss, which controlled the early stopping during training. The lighter dashed line represent the loss on the training set.} \label{training_loss}
\end{figure}

The training losses for each of the three models across all five folds can be seen in Figure~\ref{training_loss}. For each fold, it shows the training loss in the background with the light colored dotted lines and the validation loss in the foreground with the darker solid lines. It is clear that the pre-training and training on the Uniform Quantile models was fairly stable, with the losses for each fold converging to similar values. However, there is still some minor overfitting noticeable in the pre-training with the dotted lines below the solid lines. Additionally, almost all training was stopped before reaching epoch 50. Some further improvements on regularization could allow for the models to be trained longer. Looking at the Multivariate Gaussian models, training was very unstable, and only two models reached a loss below $-35$ and continued to overfit on the training data. This could be due to the data being split uniformly amongst the folds, or the choice of learning rate optimizer and scheduler. Stable training across all folds with the same configuration would be preferable. Furthermore, the training times for the two models were approximately 530 seconds per epoch for the Multivariate Gaussian models, and 158 seconds per epoch for the Uniform Quantile models.

The mean scores from the challenge metrics calculated against the test set are shown in Table~\ref{test_scores}. A caveat to these results is that the spectral scores are very expensive to calculate. Only the first 20 planets from the test set were used, so they might not be indicative of the true mean spectral scores. The Multivariate Gaussian model for fold 0 ($score_{posterior}=686.29$, $score_{spectral}=957.52$), and the ensemble of the two best Multivariate Gaussian models ($score_{posterior}=690.71$, $score_{spectral}=944.32$) showed promising results, following the work from~\cite{Ard_vol_Mart_nez_2022} and~\cite{cobb_2019}. The five Uniform Quantile models ensembled together obtained a particularly high posterior score ($score_{posterior}=696.43$), and highlights the potential "diversity"~\cite{BROWN20055} created from the K-fold training procedure. Figure~\ref{corner_plot} demonstrates a sample corner plot~\cite{corner} of the posterior distributions for one planet from the final ensemble. When submitting results for the challenge, the models were trained against the entirety of the data and all of the Uniform Quantile models and only one Multivariate Gaussian model were used in generating the predicted posterior distributions. In this case only one Multivariate Gaussian model displayed stable training. The motivation for this model selection was to capture the strength in the posterior scores from the Uniform Quantile models and the strength in the spectral scores from the Multivariate Gaussian model. Table~\ref{final_leaderboard} shows the results of this combination ranking third in the challenge with a posterior score of 623.21 and a spectral score of 915.02.

Finally, an advantage of machine learning techniques is the computational efficiency when compared with classical techniques. Sampling based on the outputs of the models was especially performant\footnote{Computer system: Windows 10, RTX 3080 Ti, i7-10700K, 32GB RAM}. For the Multivariate Gaussian models 4500 samples were drawn for each of the 665 planets in the test set, which took approximately 1.4 seconds per model including the time to load the weights. With the Uniform Quantile models, it took approximately 0.97 seconds to draw 3500 samples for each of the 665 test planets, again including the time to load the model weights each time. These times are the mean across the 5 folds per model setup.

\begin{table}
\centering
\caption{Test scores. MG denotes the Multivariate Gaussian models, while UQ denotes the Uniform Quantile models. When models were ensembled, an equal number of samples were drawn from the distributions predicted by each model.}
\label{test_scores}
\begin{tabular}{llccc}
\hline
Model & Samples & Posterior Score & Spectral Score & Final Score \\
 \hline
MG fold 0 & 2500 & \textbf{686.29} & \textbf{957.52} & \textbf{740.54} \\
MG fold 1 & 2500 & 568.06 & 910.98 & 636.64 \\
MG fold 2 & 2500 & 488.01 & 904.52 & 571.31 \\
MG fold 3 & 2500 & 483.55 & 878.53 & 562.55 \\
MG fold 4 & 2500 & 632.16 & \textbf{947.81} & 695.29 \\
MG all folds & 2500 & 615.02 & 919.52 & 675.92 \\
MG best 2 folds & 2000 & \textbf{690.71} & \textbf{944.32} & \textbf{741.43} \\
UQ fold 0 & 2500 & 662.75 & 866.93 & 703.59 \\
UQ fold 1 & 2500 & 660.18 & 896.26 & 707.39 \\
UQ fold 2 & 2500 & 662.99 & 899.09 & 710.21 \\
UQ fold 3 & 2500 & 660.05 & 897.61 & 707.56 \\
UQ fold 4 & 2500 & 653.36 & 895.30 & 701.75 \\
UQ all folds & 2500 & \textbf{696.43} & 880.15 & \textbf{733.17} \\
Top 2 MG, all UQ & 3500 & \textbf{696.49} & 880.46 & \textbf{733.28} \\
\hline
\end{tabular}
\end{table}

\section{Conclusion}
The procedure outlined in this paper achieved a final score of 681.57, ranking third overall in the 2023 Ariel Data Challenge, highlighting the strength of prior work with multivariate Gaussian distributions~\cite{cobb_2019,Ard_vol_Mart_nez_2022} and the effectiveness of an ensemble of uniform distributions. The results from the private leaderboard are shown in Table~\ref{final_leaderboard}. Based on the distribution of final scores and even the differences between the top 4, this is a challenging open problem. A few suggestions for future work include hyperparameter optimization, separate backbones, alternate feature engineering and feature selection, transfer learning with a pre-trained unsupervised learning scheme such as an encoder-decoder network, and combining the output distributions as part of the learning process such as with a Mixture of Experts~\cite{chen2022understanding}.

\begin{table}[]
\centering
\caption{Private leaderboard final ranking}
\label{final_leaderboard}
\begin{tabular}{ccccc}
\hline
Participant       & Posterior Score & Spectral Score & Final Score & Final Rank \\
\hline
Mayeul\_Aubin   & 667.39         & 895.84        & 713.08      & 1          \\
gators          & 660.87         & 871.96        & 703.09      & 2          \\
asweet          & 623.21         & 915.02        & 681.57      & 3          \\
Les3Stagios     & 613.68         & 940.49        & 679.05      & 4          \\
aescalantelopez & 602.10         & 860.43        & 653.77      & 5          \\
ofaucoz         & 583.15         & 896.55        & 645.83      & 6          \\
hieucao         & 526.28         & 916.92        & 604.41      & 7          \\
MALTO           & 530.04         & 853.33        & 594.70      & 8         \\
\hline
\end{tabular}
\end{table}

\subsubsection{Acknowledgements} I would like to thank the organizers of this challenge for hosting it and for the data and resources they have compiled. I would also like to thank the other competitors for (indirectly) pushing me to continue learning and improving my own score, and everyone who read and provided feedback on this manuscript.

\appendix\section{Appendix}\label{model.appendix}
The appendix contains additional figures for further clarification of the models, and a sample plot of the posterior distributions generated from the final ensemble compared with the ground truth trace data. In the graphs of the model architectures, m represents the batch size.

\begin{figure}[ht!]
\centering
\includegraphics[height=0.425\textheight]{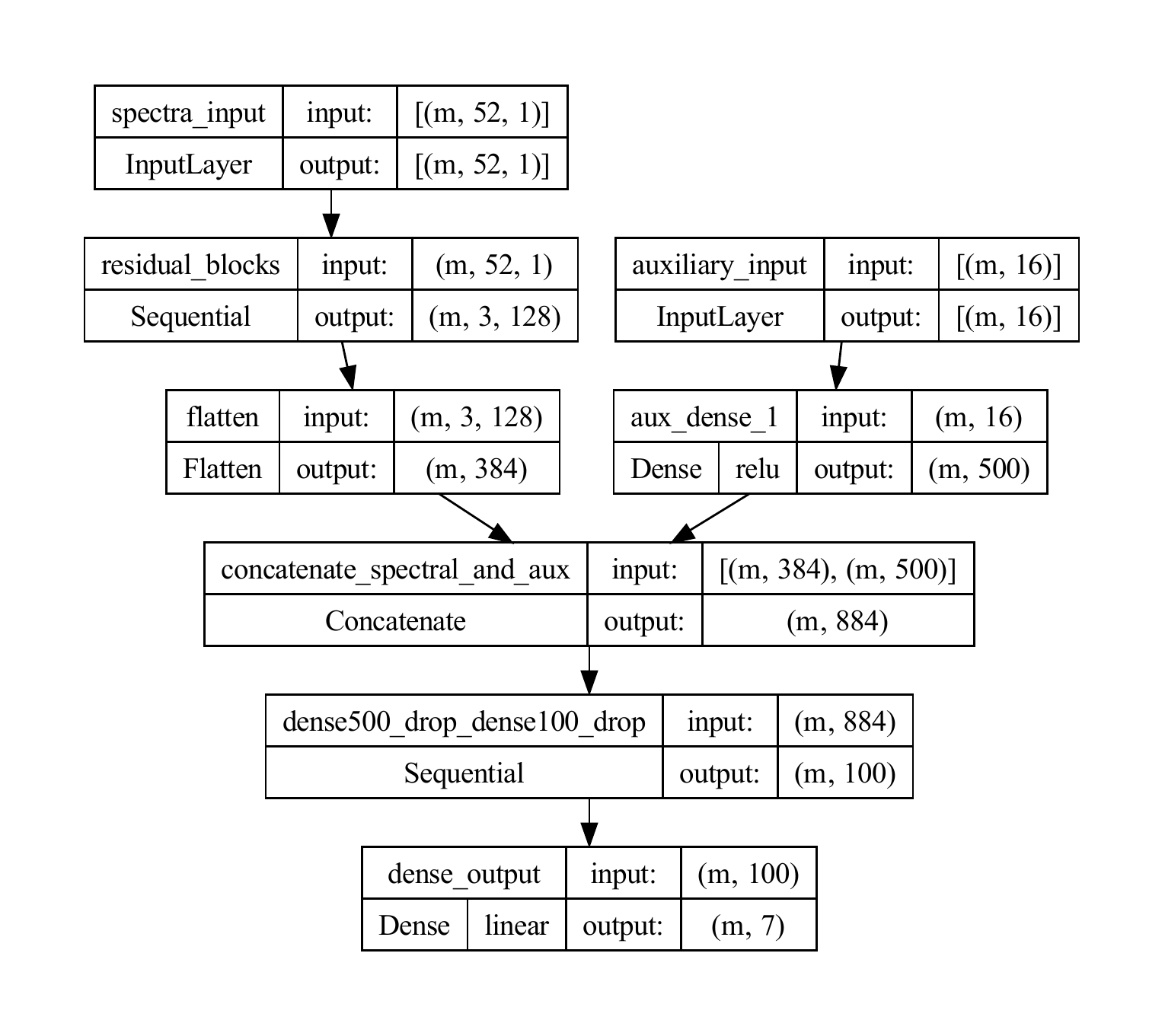}
\caption{Pretraining model.} \label{phase1model}
\end{figure}

\begin{figure}[ht!]
\centering
\includegraphics[height=0.425\textheight]{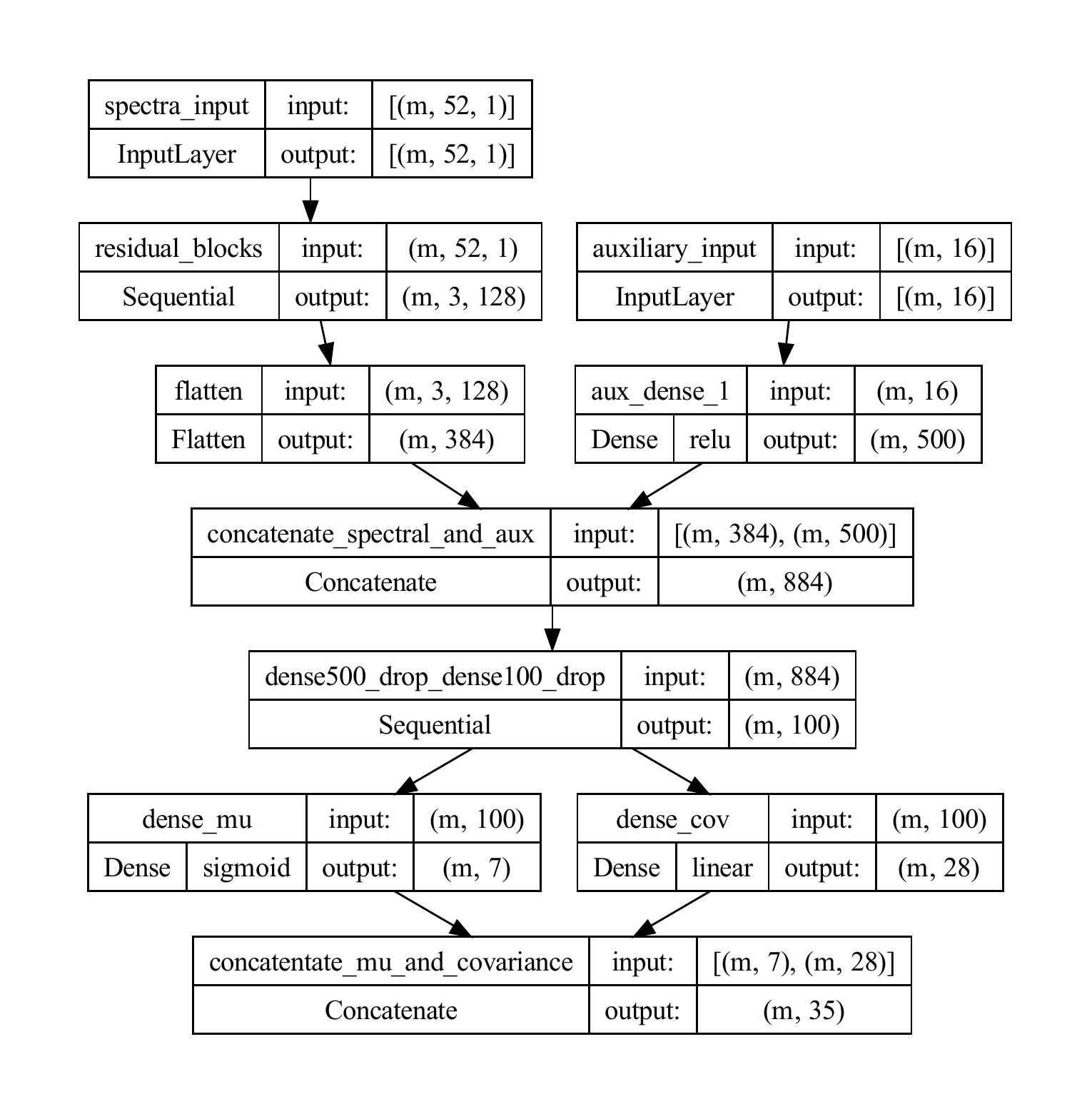}
\caption{Multivariate Gaussian model. Note that the final two dense layers were concatenated for convenience when calculating the loss.} \label{phase2MGmodel}
\end{figure}

\begin{figure}[ht!]
\centering
\includegraphics[height=0.6\textheight]{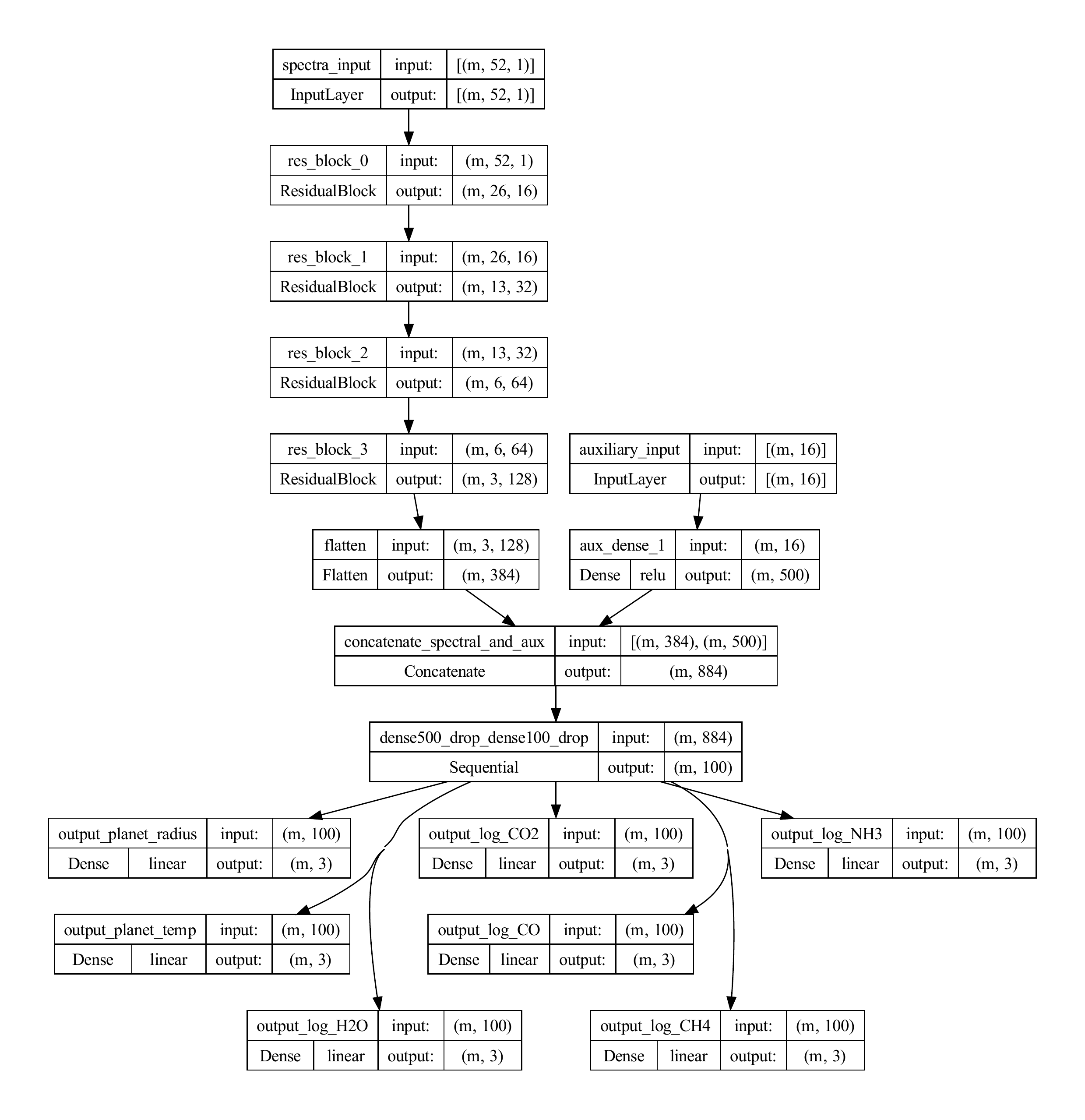}
\caption{Uniform Quantile model. There are 7 output layers for this model.} \label{phase2UQmodel}
\end{figure}

\begin{figure}[ht!]
\centering
\includegraphics[width=\textwidth]{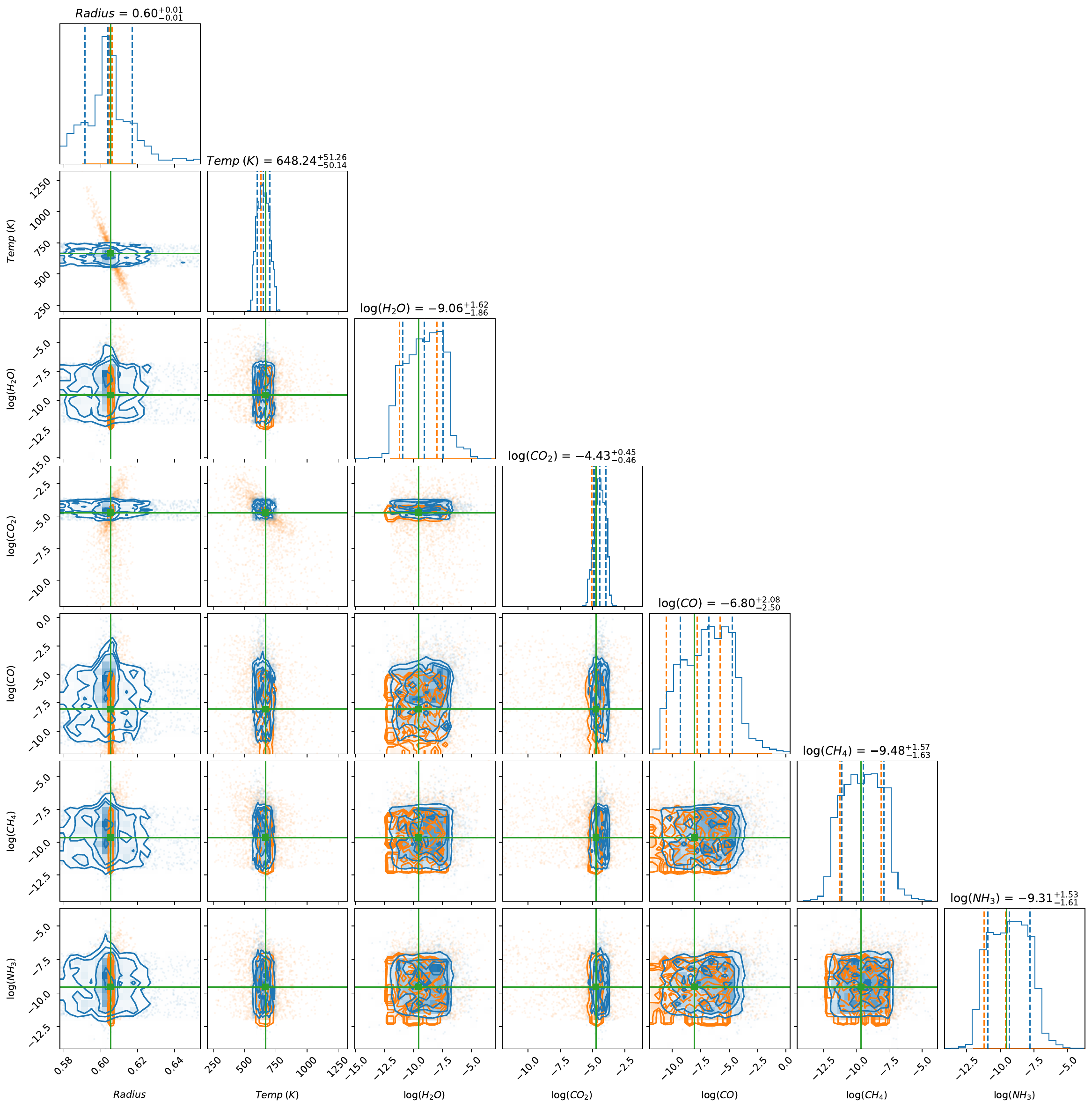}
\caption{Sample corner plot~\cite{corner} of the posterior distributions for one planet. Orange denotes the ground truth provided in the trace data, blue represents the posterior distribution sampled from the ensemble of the two best Multivariate Gaussian models and the five Uniform Quantile models, and the green lines are the weighted averages of the trace data for each target.} \label{corner_plot}
\end{figure}

\clearpage
\newpage

%
%
%
\bibliographystyle{splncs04}
\bibliography{references}

\end{document}